\documentclass[conference]{IEEEtran}

\usepackage{amssymb}
\usepackage{amsmath}
\usepackage{pgfplots}
\usepackage{graphicx}
\usepackage{float}

\begin{document}

\title{MPC Validation and Aggregation of Unit Vectors
	}

\author{\IEEEauthorblockN{Dylan Gray~~~~Joshua Joy~~~~Mario Gerla}}


%


\maketitle

\begin{abstract}
When dealing with privatized data, it is important to be able to protect against malformed user inputs. This becomes difficult in MPC systems as each server should not contain enough information to know what values any user has submitted. In this paper, we implement an MPC technique to verify blinded user inputs are unit vectors. In addition, we introduce a BGW circuit which can securely aggregate the blinded inputs while only releasing the result when it is above a public threshold. These distributed techniques take as input a unit vector. While this initially seems limiting compared to real number input, it is quite powerful for cases such as selecting from a list of options, indicating a location from a set of possibilities, or any system which uses one-hot encoding. \\
\end{abstract}

\IEEEpeerreviewmaketitle

\section{Introduction}
\begin{figure}
\includegraphics[width=3.5in]{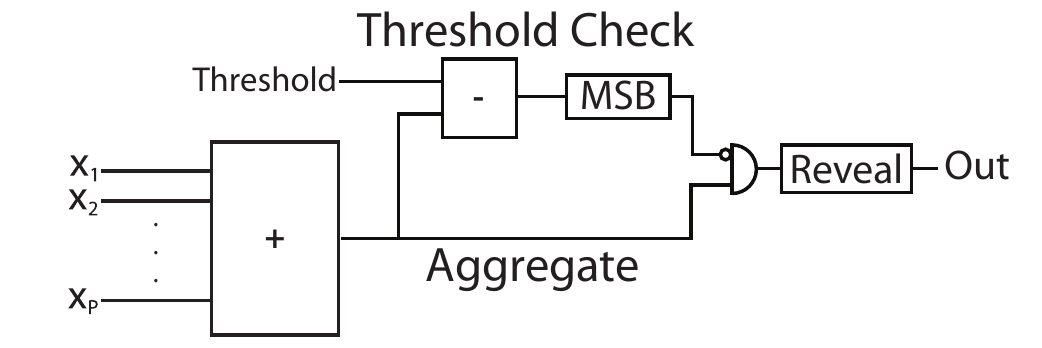}\\
\caption{BGW Multiparty Aggregation Circuit}
\label{fig:bgw_circuit}
\medskip
\small
A multiparty computation circuit which uses BGW and Shamir's Secret Sharing Algorithm to compute the aggreagate of all inputs $x_i$ and only release it if it is above the publicly known threshold.
\end{figure}

In our system, a user would like to input an N long unit vector; however, no strict subset of the servers should be able to recreate user inputs. The servers should then be able to verify that the user's input was a unit vector. In addition, the servers should then be able to aggregate all user inputs and release the value only if it is above a public threshold. The verification circuit is an implementation of the MPC protocol described in~\cite{Boyle:2016:FSS:2976749.2978429}. This scheme takes input vectors, blinds (randomizes) the values, and then splits each value across all verification servers. The result is that no strict subset of the servers which collaborate can distinguish the inputs from random. However, all servers working together can verify that an input is a unit vector without revealing any other information about the input. Finally, the servers can combine all user inputs to perform an aggregate on each element of the inputs by using the BGW circuit shown in Figure \ref{fig:bgw_circuit}. In other words, let N be the input length and $\ell$ be the number of user inputs. This data can be represented as an $\ell$xN matrix where each row is a user input, and the aggregation is a sum on each column, thus giving N aggregation results. We will now discuss the implementation of the verification and aggregation circuits as well as their performance.

\section{Verification}
Assume a user wants to submit an N long vector $\textit{V} = [v_1,~\dots,~v_N]$, and we want to verify that it is a unit vector using P parties without any party learning the user's vector input. The verification scheme is as follows:\\

\begin{enumerate}
	\item The user splits $\textit{V}$ into a PxN matrix ($Split$), such that $\sum_i \textit{Split}_{ij} = v_j$.\\
	\item The user blinds $\textit{Split}$ into a PxP matrix $\textit{Blind}$ using one of the three algorithms discussed in Section \ref{ssec:blind}.\\
	\item The user sends one Px1 column vector\\
	\begin{equation*}
	\mathit{share_i} = \begin{bmatrix}
		share_{i,1} \\
		\vdots \\
		share_{i,P}
	\end{bmatrix}
	= \begin{bmatrix}
		blind_{1,i} \\
		\vdots \\
		blind_{P,i}
	\end{bmatrix}
	\end{equation*}
	from $\textit{Blind}$ to each verifier.\\
	\item Verifiers perform an MPC sum on each row of the shares to get the Px1 vector
	\begin{equation*}
		Sums =
		\begin{bmatrix}
			sum_1\\
			\vdots\\
			sum_P 
		\end{bmatrix}
	\end{equation*}
	where
	\begin{equation*}
		sum_j = \sum_i share_{i,j}
	\end{equation*}
	\item Validation functions are performed on $\textit{Sums}$ to check $\textit{V}$ was was a unit vector.\\
\end{enumerate}

It is important to note that all operations are done in a finite field with a prime size which is randomly chosen. The recommended size is at least 256 bits, and the performance impacts of differnet field sizes are discussed in Section \ref{ssec:ver_perf}.\\

\subsection{Splitting}
The user starts with a unit vector $\textit{V} = [v_1,~\dots,~v_N]$. For each $v_j$, $j \in \{1,~\dots,~N\}$ the user generates random values $split_{i,j}$, $i \in  \{1,~\dots,~P-1\}$. Then they set $split_{p,j} = v_j - \sum_i split_{i,j}$. After this is done for all N rows, the result is \textit{Split}, a PxN matrix where $\sum_i \textit{Split}_{i,j} = v_j$.\\

\subsection{Blinding} \label{ssec:blind}
There are 3 functions defined in [1] for blinding which we implemented: $Square$, $Product$, and $Inverse$. Let $\mathcal{L}$ be a PxN matrix used to blind $\textit{Split}$, and $\textit{r}$ is a PxN matrix of random values. After the $\mathcal{L}$ matrix is generated, we perform \textit{Blind} = $\mathcal{L}$*\textit{Split} to calculate our blinded PxP matrix. It is important to note that all calculations are done in a finite field of prime size. The constructions of the $\mathcal{L}$ matrix for the various blinding techniques are below:\\

\begin{flushleft}
$Square$:\\
\end{flushleft}
\begin{equation*}
\begin{split}
\mathcal{L} &\in \textbf{F}^{PxN}\\
\mathcal{L}_{1,j} &= \textit{r}_{1,j} \text{ } \forall j\text{ s.t. }0 < j \leq N\\
\mathcal{L}_{i,j} &= \textit{r}_{1,j}^i \text{ } \forall i, j \text{ s.t. } 1 < i \leq P\text{, }0 < j \leq N\\
\end{split}
\end{equation*}

\begin{flushleft}
The result is a matrix\\
\end{flushleft}

\begin{equation*}
\mathcal{L}_{sq} = 
\begin{bmatrix}
r_{1,1} & r_{1,2} & \dots & r_{1,N}\\
r_{1,1}^2 & r_{1,2}^2 & \dots & r_{1,N}^2 \\
\vdots &  &  & \vdots\\
r_{1,1}^P & r_{1,2}^P & \dots & r_{1,N}^P\\
\end{bmatrix}
\end{equation*}
\hfill\break

\begin{flushleft}
$Product$:\\
\end{flushleft}
\begin{equation*}
\begin{split}
\mathcal{L} &\in \textbf{F}^{PxN}\\
\mathcal{L}_{i,j} &= \textit{r}_{i,j} \text{ } \forall i, j \text{ s.t. } 0 < i < P \text{, } 0 < j \leq N\\
\mathcal{L}_{P,j} &= \prod_{i = 1}^{P - 1} r_{i,j} \text{ } \forall j \text{ s.t. }\text{, }0 < j \leq N\\
\end{split}
\end{equation*}

\begin{flushleft}
The result is a matrix \\
\end{flushleft}

\begin{equation*}
\mathcal{L}_{prod} = 
\begin{bmatrix}
r_{1,1} & r_{1,2} & \dots & r_{1,N}\\
r_{2,1} & r_{2,2} & \dots & r_{2,N} \\
\vdots &  &  & \vdots\\
\prod_{i}r_{i,1} & \prod_{i}r_{i,2} & \dots & \prod_{i}r_{i,N}\\
\end{bmatrix}
\end{equation*}
\hfill\break

\begin{flushleft}
$Inverse$:\\
\end{flushleft}
\begin{equation*}
\begin{split}
\mathcal{L} &\in \textbf{F}^{PxN}\\
\mathcal{L}_{i,j} &= \textit{r}_{i,j} \text{ } \forall i, j \text{ s.t. } 0 < i < P \text{, } 0 < j \leq N\\
\mathcal{L}_{P,j} &= \Bigg[\prod_{i = 1}^{P - 1} r_{i,j}\Bigg]^{-1} \text{ } \forall j \text{ s.t. }\text{, }0 < j \leq N\\
\end{split}
\end{equation*}

\begin{flushleft}
The result is a matrix \\
\end{flushleft}
\begin{equation*}
\mathcal{L}_{inv} = 
\begin{bmatrix}
r_{1,1} & r_{1,2} & \dots & r_{1,N}\\
r_{2,1} & r_{2,2} & \dots & r_{2,N} \\
\vdots &  &  & \vdots\\
\big[\prod_{i}r_{i,1}\big]^{-1} & \big[\prod_{i}r_{i,2}\big]^{-1} & \dots & \big[\prod_{i}r_{i,N}\big]^{-1}\\
\end{bmatrix}
\end{equation*}
\hfill\break

\subsection{Sharing}
After the user calculates $\textit{Blind}$, they split it into P, Px1 column vectors 
\begin{equation*}
share_i =
\begin{bmatrix}
	share_{i,1} \\
	\vdots \\
	share_{i,P}
\end{bmatrix}
= \begin{bmatrix}
	blind_{1,i} \\
	\vdots \\
	blind_{P,i}
\end{bmatrix}
\end{equation*}
The client sends one $share_i$ to each verifier. Before the verifiers can check if the input vector is valid, they must first sum their shares into a Px1 vector
\begin{equation*}
	Sums =
	\begin{bmatrix}
		sum_1\\
		\vdots\\
		sum_P 
	\end{bmatrix}
\end{equation*}
where
\begin{equation*}
	sum_j = \sum_i share_{i,j}
\end{equation*}
This is acheived through any MPC summation scheme which does not reveal any information but the final value. Every verifier will calculate the same $Sums$ vector.\\

\subsection{Validation}
Based on the method chosen for blinding, there are 3 ways to verify that the original input vector was a unit vector [1]. $Sums$ is the Px1 vector which is the result of the MPC sum done by the verifiers. To verify, simply check the relation below holds for the chosen blinding method.\\

\begin{flushleft}
$Square$:\\
\end{flushleft}
\begin{equation*}
Sums_1^i = Sums_i~~~ \forall i,~0 < i \leq P\\
\end{equation*}

\begin{flushleft}
$Product$:\\
\end{flushleft}
\begin{equation*}
Sums_P = \prod_{i = 1}^{P - 1} Sums_i\\
\end{equation*}

\begin{flushleft}
$Inverse$:\\
\end{flushleft}
\begin{equation*}
\prod_{i = 1}^{P} Sums_i = 1\\
\end{equation*}
\hfill\newline

\subsection{Security}
At no point does any individual have enough information to reconstruct a user's input. In addition, after the user splits their value into P shares, any subset of those P shares appears random (as each share is a randomly generated bit-array). This ensures that as long as there is at least one honest party, security holds. However, for correctness to hold, all parties must act in an honestly-but-curious way.\\

\section{BGW Aggregation}
Once all users' inputs are validated, it is convenient to be able to aggregate the initial unit vectors. However, we wish for the servers to never be able to reconstruct any individual's input. We will now describe a BGW circuit which returns a sum per entry in the Nx1 input vectors across all users' inputs, and each entry's result is only released if it is above a publicly known threshold. In other words, all $\ell$ inputs can be thought of as an $\ell$xN matrix where each row is a user's input, and we will perform a sum on each column. Our BGW circuit can be seen in Figure~\ref{fig:bgw_circuit}.\\

 First, each user performs a sum on the blinded values they received from all users. Then, they use Shamir's Secret Sharing Algorithm~\cite{Shamir:1979:SS:359168.359176} to create shares and distribute one to each participating server. These are the inputs to the BGW circuit in Figure~\ref{fig:bgw_circuit}. Next, each server locally computes the sum of shares ($sum$) and subtracts the publicly known threshold ($thresh$), to get $dif = sum - thresh$. Next, we use the BGW bit conversion algorithm described in~\cite{Damgard:2006:USC:2180286.2180307} to get the encrypted bit representation of $dif$, called $dif_b$. We then take the most significant bit of $dif_b$ (called $msb$), calculate $(1-msb)*sum$, and reveal the result. If $msb = 0$ then $sum \geq thresh$, so we reveal the aggregate. If $msb=1$ then $sum < thresh$ so 0 is returned. This ensures that we only release the aggregate value if it is above $thresh$.\\

The BGW bit conversion described in~\cite{Damgard:2006:USC:2180286.2180307} works as follows. The algorithm takes as input an encrypted share $[a]_p$ and returns an array of encrypted shares $[a]_b$. When decrypted, $[a]_b$ is the bit representation of decrypted $[a]_p$. This is done by generating an array of shares $[b]_b$ which represent random bits. Next, we reveal c = $([a]_p - [b]_b)$ and convert it to bit representation in the clear. $[a]_b = c + [b]_b$ can now be computed. In our implementation, we do not reveal any bits as we only need the encrypted value of the MSB.\\

This circuit relies on manipulation of Shamir's Secret Shares, which we know is resilient to $M < P/3$ malicious adversaries where $P$ is the number of participating parties~\cite{DBLP:journals/joc/AsharovL17}. It also relies on the BGW Bit Conversion described in~\cite{Damgard:2006:USC:2180286.2180307} which they prove is secure against an adversary which controls $M < \lfloor (P-1)/2 \rfloor$ honest-but-curious participants.\\

\section{Performance}
\subsection{Verification} \label{ssec:ver_perf}
We will now discuss the performance of our implementations for the verification and aggregation schemes described above. Our verification scheme is lightweight and fast. Running on consumer grade hardware, we can run thousands of verifications simultaneously at roughly the same latency as running 100 queries simultaneously. In general, a verifier takes between 350 and 550 ms, and the client takes 20 to 50 ms for normal operating modes per query. There are five parameters which affect execution time which we will analyze in turn: field size, blinding algorithm, number of simultaneous queries, length of input vector, and number of verifiers.\\

\begin{figure}
\includegraphics[width=3.5in]{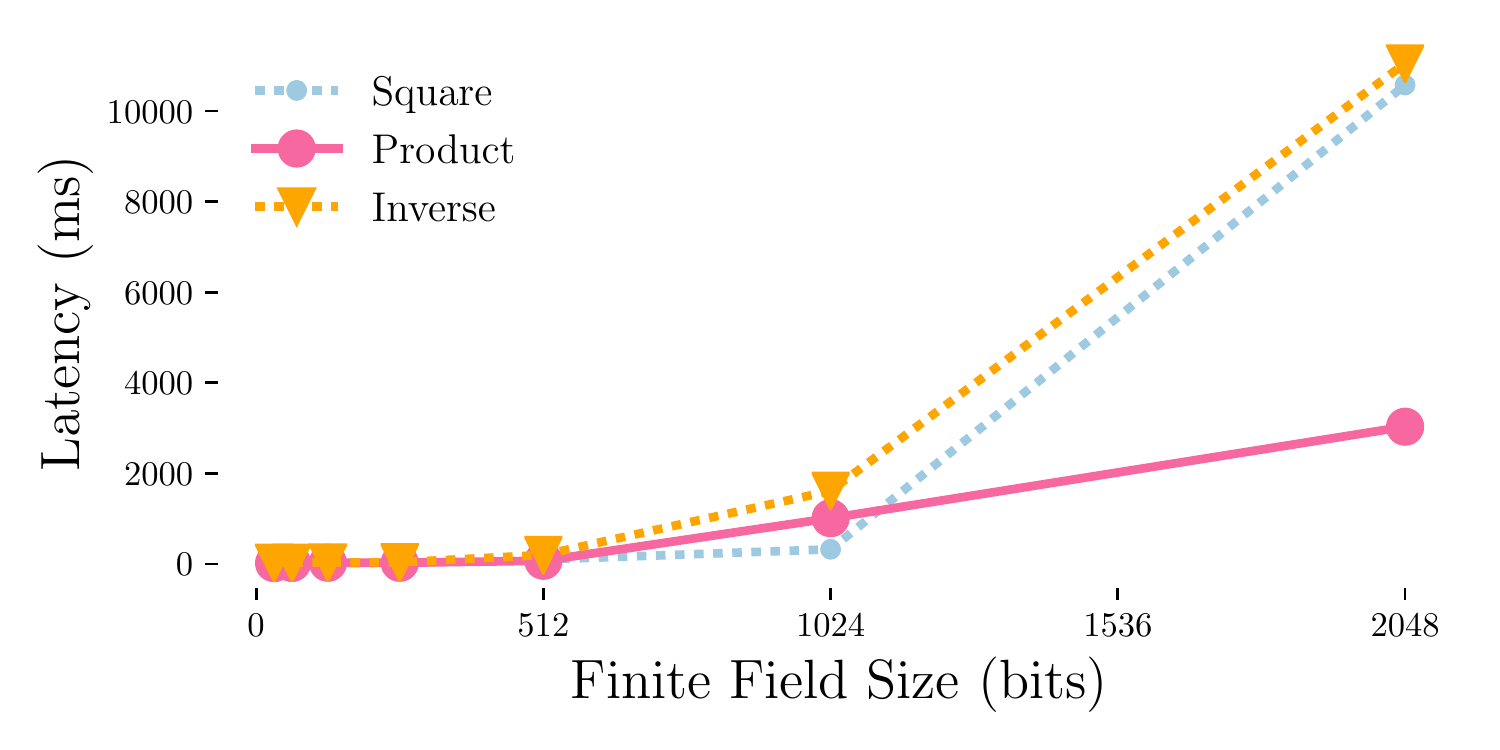}\\
\caption{Algorithm Performance (Client)}
\label{fig:alg_latency}
\medskip
\small
Client latency for various sizes of finite field for the 3 blinding algorithms.
\end{figure}

\begin{figure}
\includegraphics[width=3.5in]{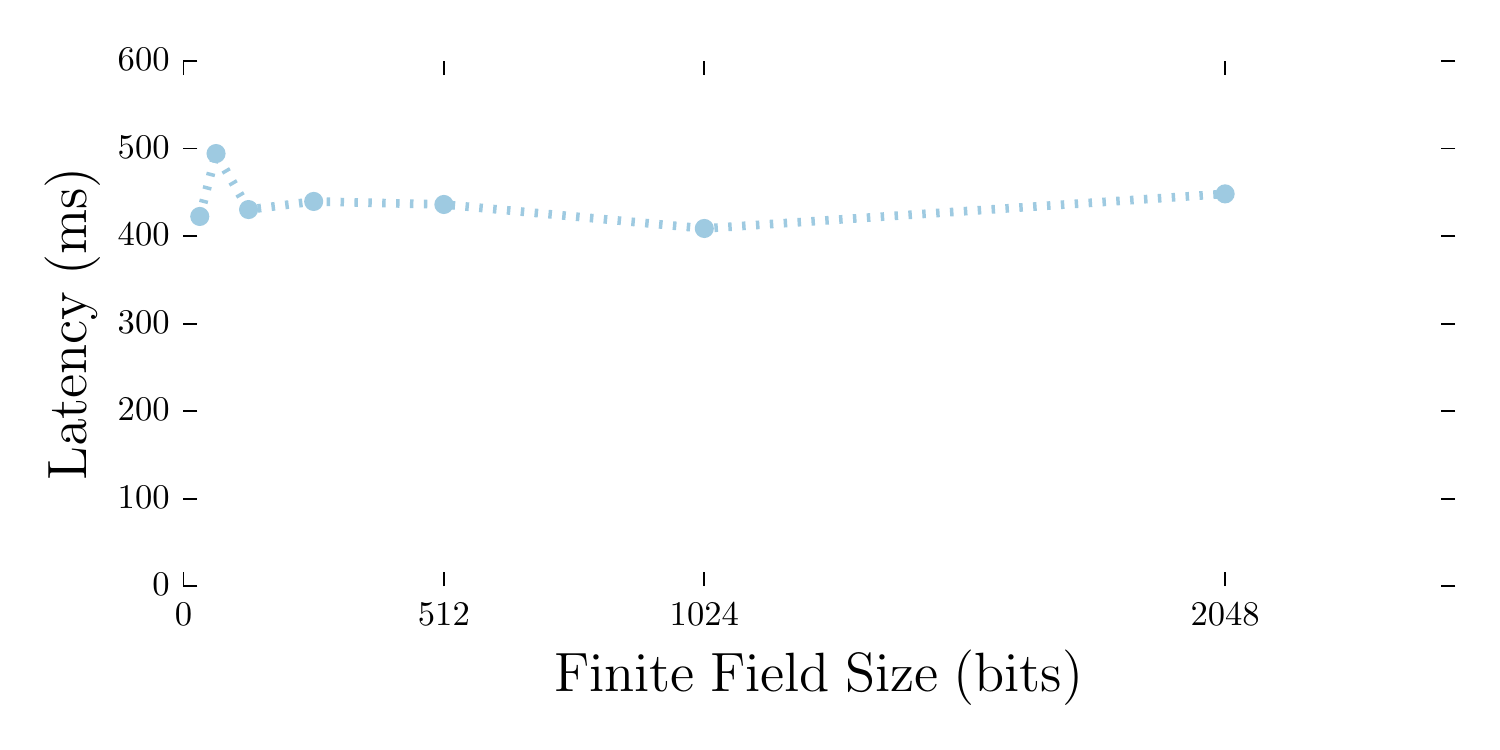}\\
\caption{Finite Field Size vs Latency (Server)}
\label{fig:ffs_server}
\medskip
\small
The latency observed by the servers for various finite field size.
\end{figure}

When verifying, a random prime is selected for the finite field size used in all our arithmetic. The length of this field does not drastically affect the verifiers, but the client's performance is drastically impacted by the number of bits in the chosen prime. The client is decently efficient up to 512 bit primes, but it slows significantly beyond that. The graphs of finite field size vs latency for the client and verifiers can be seen in Figures \ref{fig:alg_latency} and \ref{fig:ffs_server} respectively.\\


The $Square$ and $Product$ algorithms for verification have similar latencies for small finite field sizes, but $Product$ is significantly faster than $Square$ for large finite field sizes (over 1024 bits). In addition, $Inverse$ is slower for the client at all finite field sizes. This relationship can be seen in Figures \ref{fig:alg_latency} and \ref{fig:alg_table}. A mojority of the client processing time is spent blinding and transfering data over the newtork. Because of this, we see large changes in performance due to blinding algorithm choice. The $Product$ algorithm is much more efficient than the alternatives because it only needs to execute $(P-1)$ multiplications per value to blind, whereas $Square$ performs $(P-1)$ exponentiation operations, and $Inverse$ performs $(P-1)$ multiplications and one exponentiation. The majority of server time is not spent doing the final verification step, so we do not see a change in run-time for the servers when we change the algorithm used.\\

\begin{figure}
\begin{center}
\begin{tabular}{|c|c|c|}
\hline
Mode & Client (ms) & Server (ms) \\
\hline
Square & 29.75.5 & 436.8968 \\
Product & 29.65 & 562.78 \\
Inverse & 45.24 & 473.94 \\
\hline
\end{tabular}\\
\caption{Algorithm vs Latency}
\label{fig:alg_table}
\medskip
\small
The latency for client and server with different algorithms with a finite field size of 256 bits, 3 parties, and input vectors 100 values long. All algorithms are efficient at this size of finite field.
\end{center}
\end{figure}

\begin{figure}
\includegraphics[width=3.5in]{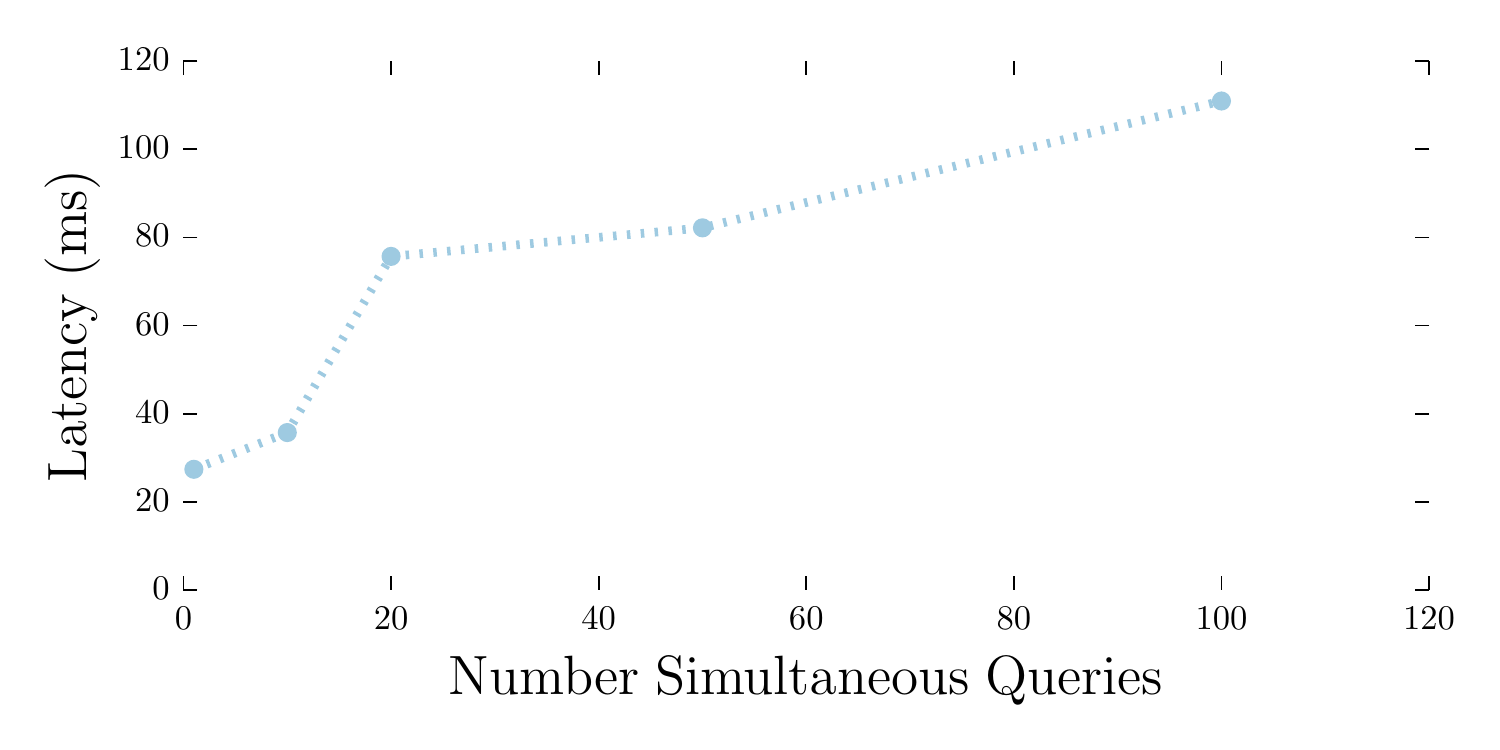}\\
\caption{Number Queries vs Latency (Client)}
\label{fig:num_client}
\medskip
\small
Latency observed by client while sending many simultaneous queries for verification.
\end{figure}

\begin{figure}
\includegraphics[width=3.5in]{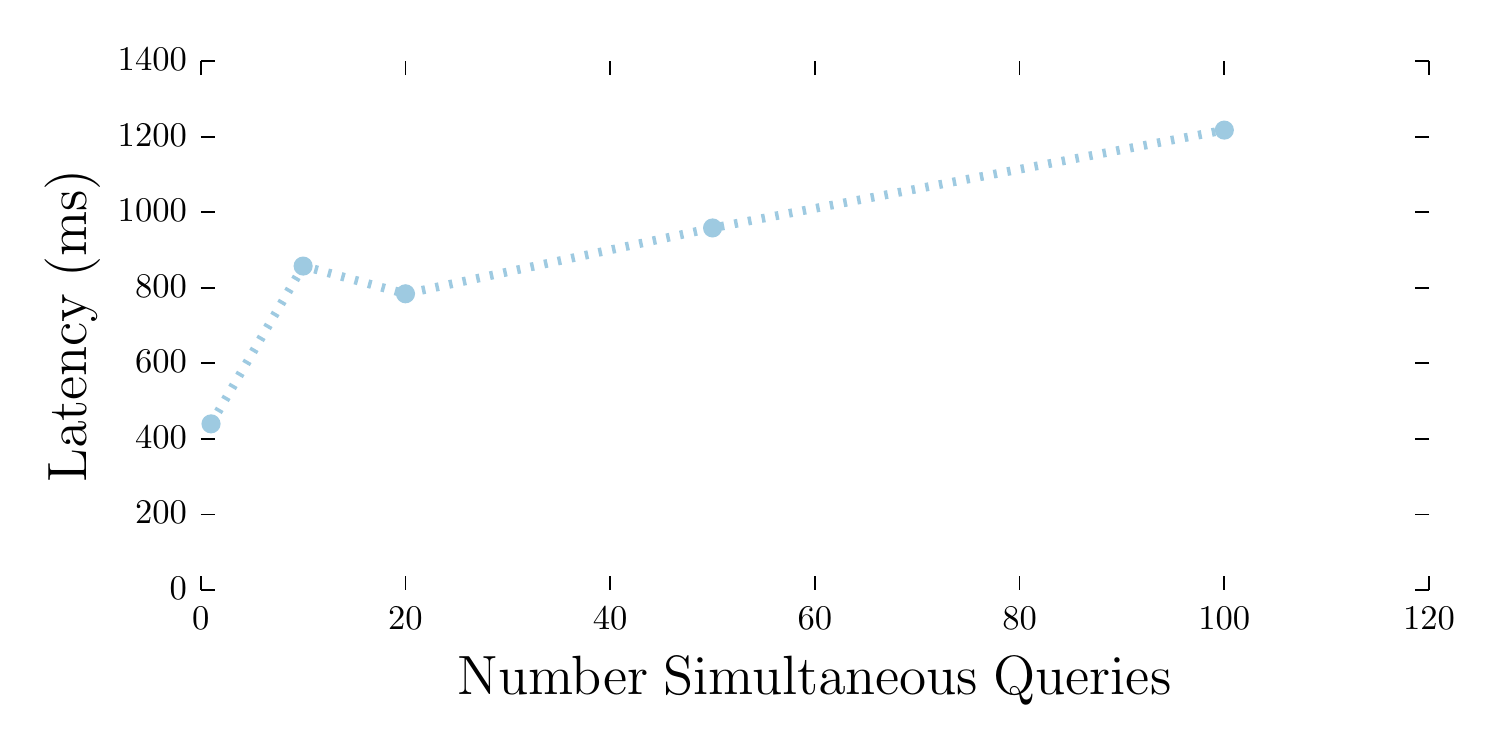}\\
\caption{Number Queries vs Latency (Server)}
\label{fig:num_server}
\medskip
\small
Server latency when processing multiple simultaneous verification requests.
\end{figure}

The number of simultaneous queries, unsurprisingly, affects latency. This linear slowdown is observed by both clients and verifiers. However, the slowdown has a gradual slope. Increasing from 1 query to 100 simultaneous queries only causes a slowdown of roughly 4 times. Figures \ref{fig:num_client} and \ref{fig:num_server} show the relationship between the number of simultaneous queries and latency for clients and verifiers respectively.\\


\begin{figure}
\includegraphics[width=3.5in]{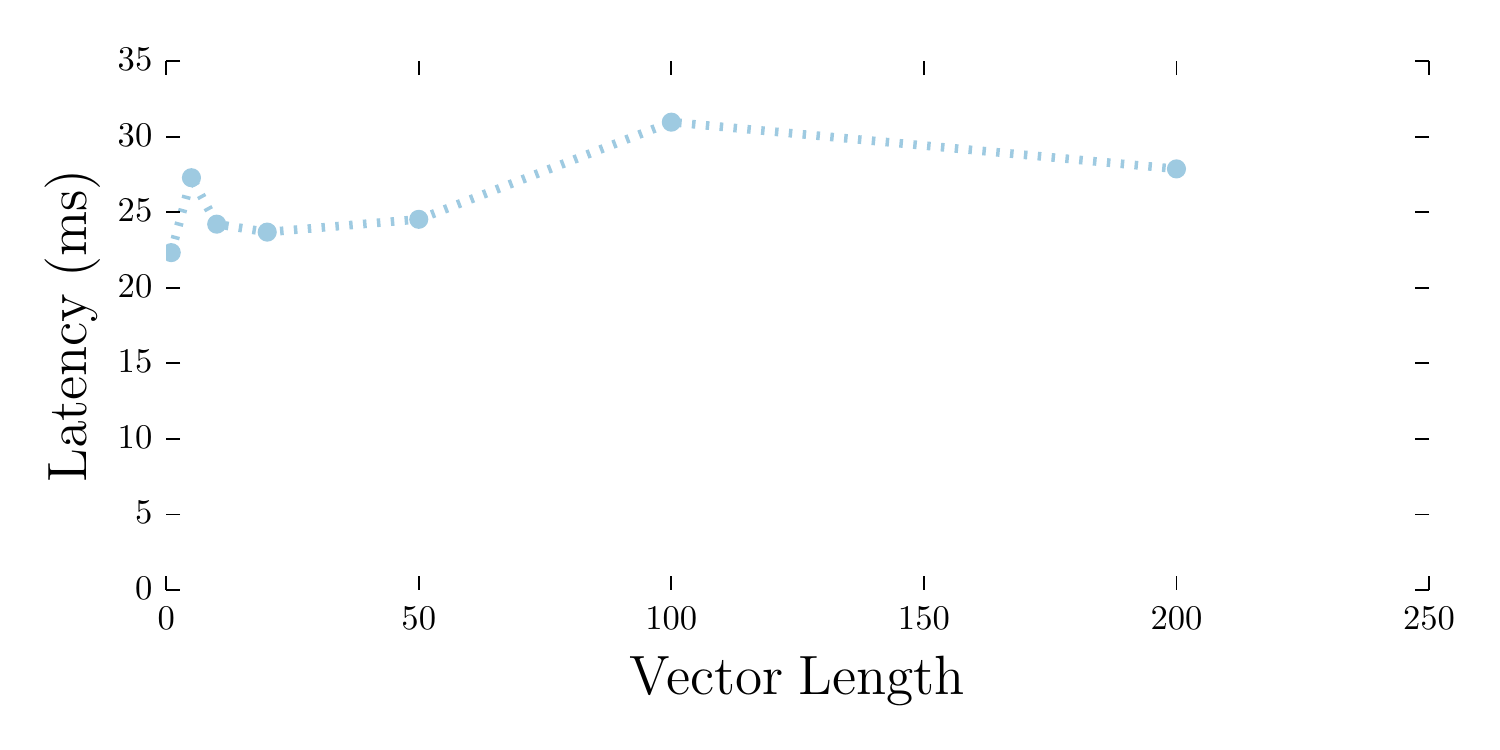}\\
\caption{Vector Length vs Latency (Client)}
\label{fig:length_client}
\medskip
\small
The latency observed by clients for various sizes on input vector.
\end{figure}

\begin{figure}
\includegraphics[width=3.5in]{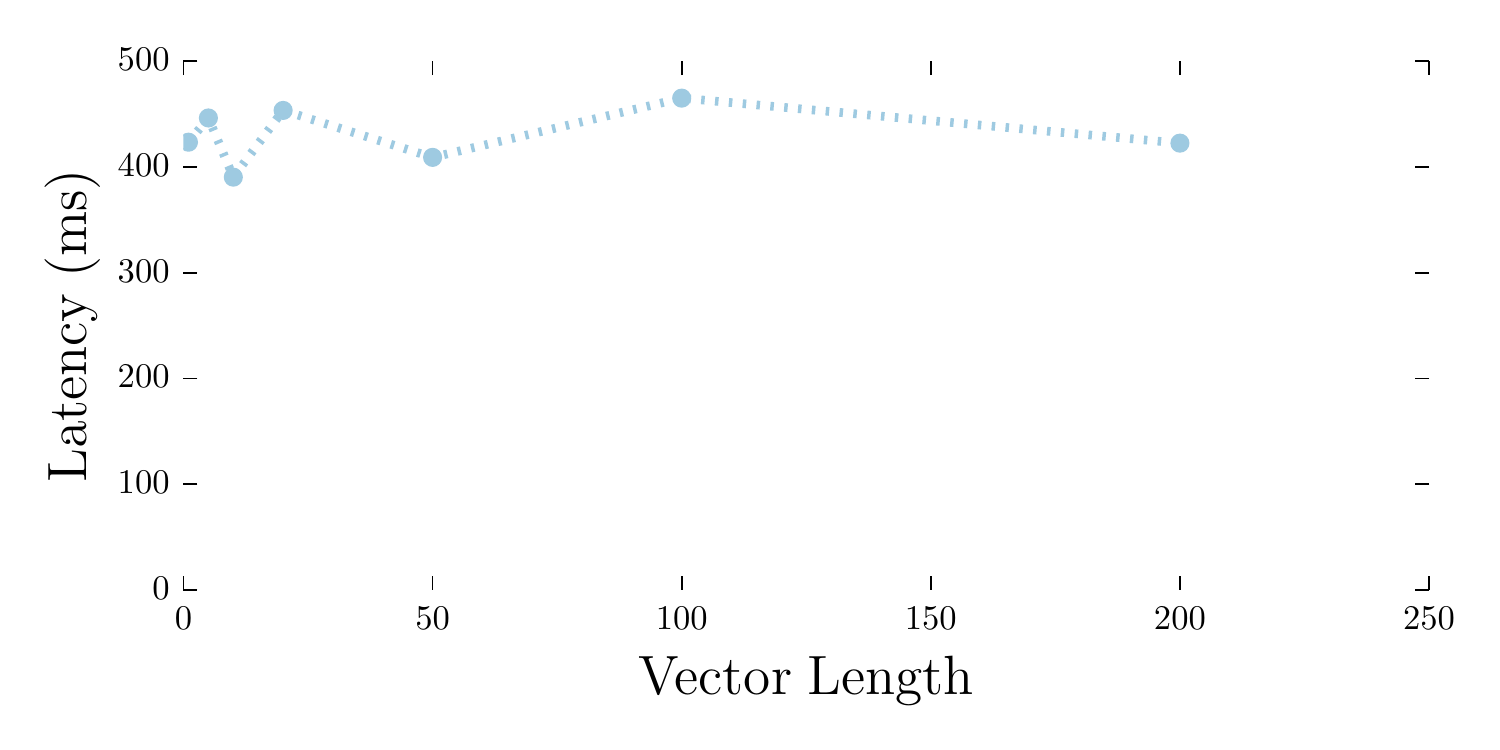}\\
\caption{Vector Length vs Latency (Server)}
\label{fig:length_server}
\medskip
\small
Server latency for various input vector lengths.
\end{figure}

Our verification algorithm is highly parallelizable. Each element in the original input vector can be handled separately, and the only instance where threads must be joined is on the final check performed in the $\textit{validation}$ step. This leads to vector size not impacting latency for the clients or verifiers. All differences in timings can be attributed to external factors, such as random generation of larger numbers or an increase in network load due to other users on the network. The relationship between vector length and latency for the client and verifiers is shown in Figures \ref{fig:length_client} and \ref{fig:length_server}.\\


\begin{figure}
\includegraphics[width=3.5in]{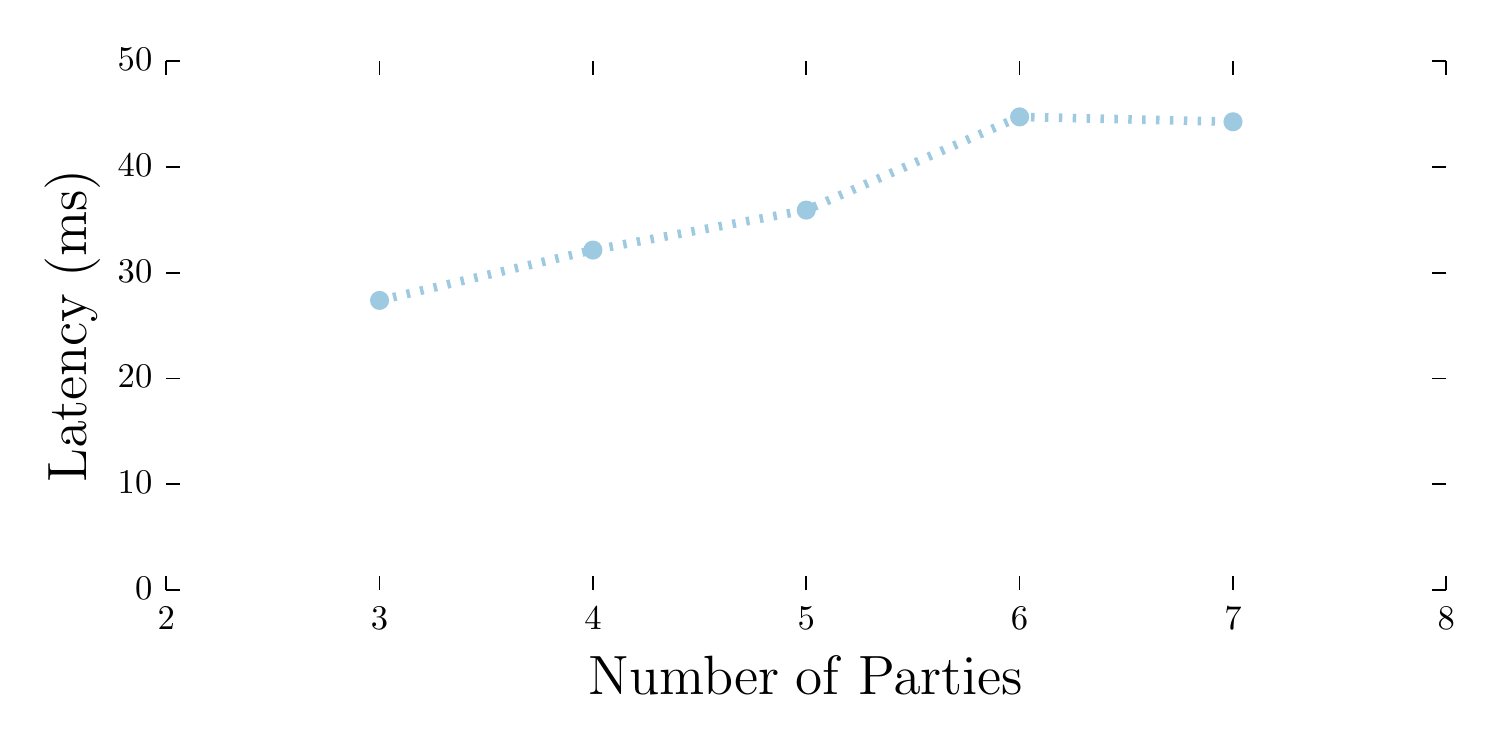}\\
\caption{Number of Parties vs Latency (Client)}
\label{fig:parties_client}
\medskip
\small
Client latency vs number of parties used for verification.
\end{figure}

\begin{figure}
\includegraphics[width=3.5in]{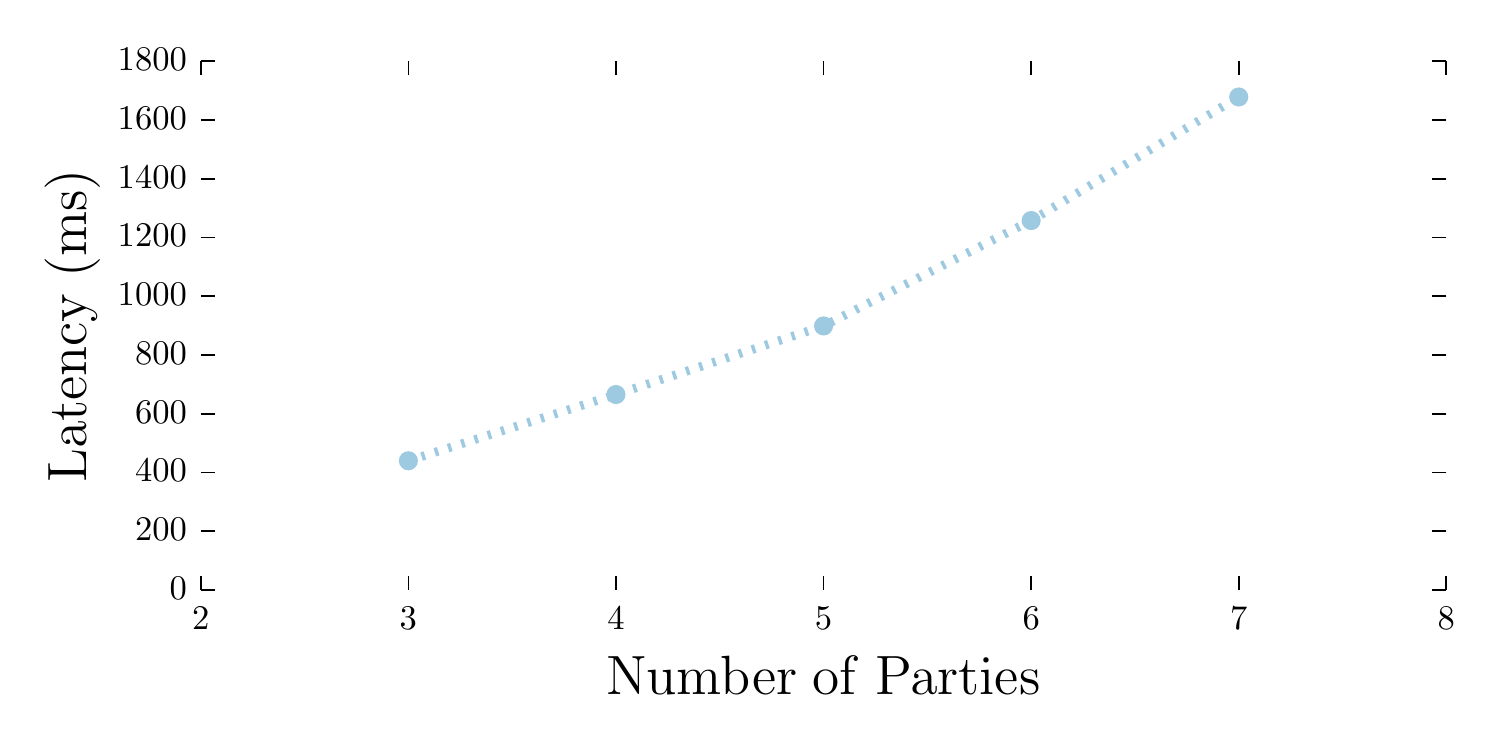}\\
\caption{Number of Parties vs Latency (Server)}
\label{fig:parties_server}
\medskip
\small
Latency seen by servers for various number of parties used in verification process.
\end{figure}

During our verification process, every verification party communicates with all other servers, meaning our communication complexity is O($P^2$) where $P$ is the number of verification parties. This relationship is seen in our experimental results for the verifiers. However, the client's dependency is linear in P because an increased number of parties linearly relates to the number of blinded shares to create. These relationships can be seen in Figures \ref{fig:parties_client} and \ref{fig:parties_server}.\\


\subsection{BGW Aggregation}
BGW bit conversion is incredibly expensive. As shown in~\cite{Damgard:2006:USC:2180286.2180307}, the BGW bit conversion protocol requires $O(L \log L)$ BGW multiplications where $L$ is the nubmer of bits we wish to convert to. Each of these multiplications involves every server sending and receiving data to and from every other server, leading to large delays. Our findings mirrored these results. Once parallelized, our consumer-level hardware performed bit conversion in the following times:\\

\begin{center}
\begin{tabular}{|c|c|}
\hline
$L$ (bits) & Server (s)\\
\hline
8 & 49.4\\
32 & 508\\
64 & 1,850\\
\hline
\end{tabular}\\
\end{center}
\hfill\newline

This is incredibly inefficient, even though the algorithms are run in parallel. These times will improve with enterprise-level hardware, however they will still be quite slow. Alternative techniques (such as GMW circuits) are recommended.\\

\section{Conclusion}
This paper has described an implementation for the validation and aggregation of secret unit vector inputs in an MPC system. It has provided a fast, scalable system for validation with minimal overhead and computation requirements while retaining privacy as long as one honest server exists. It also provided a (slow) scheme for aggregation on secret inputs which only releases the sum if above a publicly known threshold.


\nocite{*}
\bibliography{report_formatted}{}

\begin{thebibliography}{1}
\providecommand{\url}[1]{#1}
\csname url@samestyle\endcsname
\providecommand{\newblock}{\relax}
\providecommand{\bibinfo}[2]{#2}
\providecommand{\BIBentrySTDinterwordspacing}{\spaceskip=0pt\relax}
\providecommand{\BIBentryALTinterwordstretchfactor}{4}
\providecommand{\BIBentryALTinterwordspacing}{\spaceskip=\fontdimen2\font plus
\BIBentryALTinterwordstretchfactor\fontdimen3\font minus
  \fontdimen4\font\relax}
\providecommand{\BIBforeignlanguage}[2]{{%
\expandafter\ifx\csname l@#1\endcsname\relax
\typeout{** WARNING: IEEEtran.bst: No hyphenation pattern has been}%
\typeout{** loaded for the language `#1'. Using the pattern for}%
\typeout{** the default language instead.}%
\else
\language=\csname l@#1\endcsname
\fi
#2}}
\providecommand{\BIBdecl}{\relax}
\BIBdecl

\bibitem{Boyle:2016:FSS:2976749.2978429}
\BIBentryALTinterwordspacing
E.~Boyle, N.~Gilboa, and Y.~Ishai, ``Function secret sharing: Improvements and
  extensions,'' in \emph{Proceedings of the 2016 ACM SIGSAC Conference on
  Computer and Communications Security}, ser. CCS '16.\hskip 1em plus 0.5em
  minus 0.4em\relax New York, NY, USA: ACM, 2016, pp. 1292--1303. [Online].
  Available: \url{http://doi.acm.org/10.1145/2976749.2978429}
\BIBentrySTDinterwordspacing

\bibitem{Shamir:1979:SS:359168.359176}
\BIBentryALTinterwordspacing
A.~Shamir, ``How to share a secret,'' \emph{Commun. ACM}, vol.~22, no.~11, pp.
  612--613, Nov. 1979. [Online]. Available:
  \url{http://doi.acm.org/10.1145/359168.359176}
\BIBentrySTDinterwordspacing

\bibitem{Damgard:2006:USC:2180286.2180307}
\BIBentryALTinterwordspacing
I.~Damg{\aa}rd, M.~Fitzi, E.~Kiltz, J.~B. Nielsen, and T.~Toft,
  ``Unconditionally secure constant-rounds multi-party computation for
  equality, comparison, bits and exponentiation,'' in \emph{Proceedings of the
  Third Conference on Theory of Cryptography}, ser. TCC'06.\hskip 1em plus
  0.5em minus 0.4em\relax Berlin, Heidelberg: Springer-Verlag, 2006, pp.
  285--304. [Online]. Available: \url{http://dx.doi.org/10.1007/11681878\_15}
\BIBentrySTDinterwordspacing

\bibitem{DBLP:journals/joc/AsharovL17}
\BIBentryALTinterwordspacing
G.~Asharov and Y.~Lindell, ``A full proof of the {BGW} protocol for perfectly
  secure multiparty computation,'' \emph{J. Cryptology}, vol.~30, no.~1, pp.
  58--151, 2017. [Online]. Available:
  \url{http://dx.doi.org/10.1007/s00145-015-9214-4}
\BIBentrySTDinterwordspacing

\end{thebibliography}
\bibliographystyle{IEEEtran}

\end{document}